\documentclass[aps,showpacs,manuscript,12pt]{revtex4}
\usepackage{amsfonts}
\usepackage{amssymb}
\usepackage{amsmath}
\usepackage{graphicx}

\begin{document}

\title{\textbf{Theory of multi-point probability densities for
incompressible Navier-Stokes fluids}}
\author{C. Asci$^{1}$ and M. Tessarotto$^{1,2}$}
\affiliation{$^{1}$Department of Mathematics and Informatics,\\
University of Trieste, Trieste, Italy \\
$^{2}$ Consortium for Magnetofluid Dynamics, Trieste, Italy }

\begin{abstract}
An open problem arising in the statistical description of turbulence is
related to the \textit{theoretical prediction based on first principles} of
the so-called multi-point velocity probability density functions (PDFs)
characterizing a Navier-Stokes fluid.

In this paper it will be shown that - based on a suitable axiomatic approach
- a solution to this problem can actually be achieved based on the so-called
inverse kinetic theory (IKT), recently developed for incompressible fluids.
More precisely, we intend to show, based on the requirement that \textit{the
Boltzmann-Shannon entropy for the s-point velocity PDF ($f_s$) is
independent of the order $s$ and is also maximal at all times}, that all
multi-point PDFs are \textit{necessarily factorized in terms of the
corresponding 1-point velocity PDF} ($f_1$). As a consequence the
multi-point PDFs usually considered for the phenomenological description of
turbulence can be theoretically predicted \textit{based on the knowledge of $%
f_1$ achieved by means of IKT}.
\end{abstract}
\date{\today }
\maketitle

\pagebreak

\section{Introduction}

In the context of the statistical description of fluids, the
problem of the determination of multi-point PDFs arises (at least)
in two circumstances:

\begin{itemize}
\item the first one occurs in the phenomenological description of turbulence
(see for example Monin and Yaglom \cite{Monin1975} 1975 and Pope, 2000 \cite%
{Pope2000}). In such a context, in fact, the statistical behavior
of fluids is often described in terms of statistical frequencies
defined for multi-point velocity spatial increments (however,
similar frequencies can be established also for other fluid
fields, such as vorticity, scalar pressure, temperature, etc.).

\item the second one is the so-called Monin-Lundgren hierarchy \cite%
{Monin1967,Lundgren1967}, based on the construction of an infinite
set of equations for suitable ensemble-averaged multi-point PDFs
(ML approach). Such a theory should provide, in principle, also a
theoretical model for the phenomenological description of
turbulence and as a consequence be able to predict also the
precise form of the velocity-difference PDF observed
experimentally in HIST (homogenous, isotropic and stationary
turbulence). The goal the ML approach is actually to predict the
time evolution of the ensemble average of the 1-point PDF, to be
defined in terms of a suitable (and yet to be defined)
ensemble-averaging operator.
\end{itemize}

Several open issues are related to the ML approach. \ These
concern, in particular, the search of possible \textit{exact
particular solutions} of the ML hierarchy represented by a
\textit{finite set} of multi-point PDFs. It is well known that the
construction of "closure conditions" of this type for the ML
hierarchy (\textit{closure problem}) remains one of the major
unsolved theoretical problems in fluid dynamics. In practice,
however, the program of constructing (\textit{exact}) theories of
this type or (in some sense) \textit{approximate}, and holding for
arbitrary fluid fields, is still open due to the difficulty of
preserving the full consistency with the fluid equations. In fact,
it is well known that many of the customary statistical models
adopted in turbulence theory - which are based on closure
conditions of various type - typically reproduce at most only in
some approximate (i.e., asymptotic) sense the fluid equations.

This leaves fundamentally unsolved the problem of the construction
of a consistent theoretical model for the multi-point PDFs arising
in the phenomenological description of turbulence.

The goal of this paper is \textit{to prove} (see THM.1 below in
Section 2) \textit{that under suitable assumptions all multi-point
velocity PDFs characterizing a turbulent NS fluid are factorizable
in terms of the corresponding 1-point velocity PDF. }

As a result (see Sec.3) the treatment of multi-point PDFs can be
reached in
the context of IKT (inverse kinetic theory \cite%
{Ellero2005,Tessarotto2006,Tessarotto2007a,Tessarotto2008z,Tessarotto2010c})
based on the 1-point velocity statistics. It follows the
fundamental consequence that the multi-point PDFs usually
considered for the phenomenological description of turbulence
\textit{can actually be theoretically predicted} in this way! In
particular, in the case of local
Gaussian 1-point PDF \cite{Tessarotto2010d} this permits to achieve \textit{%
explicit analytic representations} of the multipoint velocity PDFs
usually considered in the phenomenological description of
turbulence.

\section{Multi-point statistical model}

The description of fluids, and more generally of continua, is
based on the introduction of a suitable set of fluid fields
$\left\{ Z\right\} \equiv \left\{ Z_{i},i=1,k\right\} $ \
satisfying a closed set of PDEs denoted as \textit{fluid
equations. }In the case of a fluid obeying of the
incompressible Navier-Stokes equations (INSE, \textit{NS fluid}), they are $%
\left\{ Z\right\} \equiv \left\{ \rho
_{0},\mathbf{V},p_{1},S_{T}\right\}.$ In particular, here $\rho
_{0}$ (the mass density) and $S_{T}$ (the
thermodynamic entropy) are both assumed constant in $\overline{\Omega }%
\times I$, where the latter requirement implies that for
isentropic flows
the equation $\partial S_{T}(t)/\partial t=0$ must hold identically for all $%
t \in I$. In addition $\mathbf{V}$ and $p_{1}$ denote respectively
the fluid velocity and the kinetic pressure; in particular,
$p_{1}$ is defined as the strictly positive function
\begin{equation}
p_{1}(\mathbf{r},t)=p(\mathbf{r},t)+p_{0}(t)+\phi (\mathbf{r},t),
\label{p_1}
\end{equation}%
where $p(\mathbf{r},t),p_{0}(t)$ and $\phi (\mathbf{r},t)$
represent respectively the fluid pressure, the (strictly-positive)
pseudo-pressure and the (possible) potential associated to the
conservative volume force density acting on the fluid [see the
Appendix, Eq.(\ref{potential})].

The \textit{statistical description }usually adopted for turbulent
flows consists, instead, in the introduction of appropriate
axiomatic approaches denoted \textit{statistical models}, i.e.,
sets $\left\{ f,\Gamma \right\} $
formed by a suitable probability density function (PDF) and a phase-space $%
\Gamma $ (subset of $%
\mathbb{R}
^{n}$) on which $f$ is defined.\ \ By definition, a statistical model $%
\left\{ f,\Gamma \right\} $ realizes a statistical description of
the fluid
if it is possible to define a mapping%
\begin{equation}
\left\{ f,\Gamma \right\} \Rightarrow \left\{ Z\right\} ,
\label{relationship}
\end{equation}%
which allows the representation in terms of $f$ either:

A) of \textit{the complete set }or more generally only\textit{\ B)
of a subset} of the \textit{fluid fields} $\left\{ Z\right\}
\equiv \left\{ Z_{i},i=1,n\right\} $ which define the fluid state.

In particular, the fluid fields $Z_{i}(\mathbf{r},t)\in \left\{
Z\right\} $
are assumed as functionals of $f$ represented by suitable "velocity" \textit{%
moments }(of\textit{\ \ }$f$). \ In both cases their construction
involves,
besides the specification of the phase space ($\Gamma $) and the \textit{%
probability density function }(PDF) $f,$ the identification of the
functional class to which $f$\textit{\ }must belong, denoted as
$\left\{ f\right\} .$\ Statistical approaches fulfilling either
property A or B will
be denoted respectively \textit{complete} and \textit{incomplete} \textit{%
statistical models. \ }For definiteness in the remainder we shall
consider only complete statistical models.

'A priori' the PDF $f$ to be used in a statistical model of this
type may be identified with an $N$-point PDF of the form
\begin{equation}
f_{N}(\mathbf{x},t)\equiv
f_{N}(\mathbf{x}_{1},\mathbf{...,x}_{N},t),
\end{equation}%
and required to satisfy the normalization condition:

\begin{equation}
\int\limits_{U^{N}}\prod\limits_{j=1,N}d^{3}\mathbf{v}_{j}\ f_{N}(\mathbf{x}%
,t)=1,  \label{normalization}
\end{equation}%
i.e., to be a \textit{velocity probability density }(in the velocity space%
\textit{\ }$U^{N});$\textit{\ }moreover, $N\geq 1$ and for all $i=1,N,$ $%
\mathbf{x}_{i}=\left( \mathbf{r}_{i},\mathbf{v}_{i}\right) ,$
$\mathbf{r}_{i} $ and $\mathbf{v}_{i}$ denote $N$ position and
velocity vectors,
respectively belonging to the configuration space of the fluid $\overline{%
\Omega }$ and a suitable velocity space $U$ to be identified with $%
\mathbb{R}
^{3}.$ \ In particular, consistent with the physical requirement
of a NS
fluid [i.e. the existence of a strong solution of INSE in the set $\overline{%
\Omega }\times I$], the following assumptions are introduced for
$f\equiv f_{N}$:

\begin{itemize}
\item \textit{Axiom }\#1 (\textit{symmetry condition}): $f_{N}(\mathbf{x}%
_{1},\mathbf{...,x}_{N},t)$ is symmetric w.r. to arbitrary permutation $%
\left( \mathbf{x}_{1},\mathbf{...,x}_{N}\right) ,$ i.e.,
satisfying the
invariance condition%
\begin{equation}
f_{N}(\left( \mathbf{x}_{1},\mathbf{...,x}_{N}\right)
,t)=f_{N}(\left( \mathbf{x}_{1},\mathbf{...,x}_{N}\right) ^{\prime
},t);  \label{INVARIANCE}
\end{equation}

\item \textit{Axiom }\#2 (\textit{reduced }$s$\textit{-body PDFs}): $f_{N}(%
\mathbf{x}_{1},\mathbf{...,x}_{N},t)$ defines for all $s=1,N-1$ the \textit{%
reduced }$s$\textit{-body PDFs}%
\begin{equation}
f_{s}(\mathbf{x}_{1},\mathbf{...,x}_{s},t)=\frac{1}{\mu (\Omega )}%
\int\limits_{\Omega }d^{3}\mathbf{r}_{s+1}\int\limits_{U}d^{3}\mathbf{v}%
_{s+1}\ f_{s+1}(\mathbf{x}_{1},\mathbf{...,x}_{s+1},t),
\label{reduced s-body PDF}
\end{equation}%
where $\mu (\Omega )=\int\limits_{\Omega }d^{3}\mathbf{r}_{1}$ is
assumed
finite and $>0$. Hence each $f_{s}$ satisfies, thanks to (\ref{normalization}%
), the normalization
\begin{equation}
\int\limits_{U^{s}}\prod\limits_{j=1,s}d^{3}\mathbf{v}_{j}\ f_{s}(\mathbf{x}%
_{1},\mathbf{...,x}_{s},t)=1;  \label{normalization-s}
\end{equation}

\item \textit{Axiom }\#3 (\textit{fluid moments}): $f_{N}(\mathbf{x}_{1},%
\mathbf{...,x}_{N},t)$ determines uniquely the local fluid fields.
Thus,
introducing suitable weight functions $G_{i}(\mathbf{r}_{k},\mathbf{v}_{k}%
\mathbf{,}t),$ for all $k=1,n$ the local fluid fields\ $Z_{i}(\mathbf{r}_{k}%
\mathbf{,}t)$ to be identified with $\mathbf{V},p_{1}$ [both
evaluated at
the local position $\mathbf{r}_{k}$ and time $t$ belonging to $\overline{%
\Omega }\times I$] are taken of the form:
\begin{equation}
\left.
\begin{array}{c}
\frac{1}{\mu (\Omega )^{N-1}}\int_{\Omega
^{N-1}}\prod\limits_{h=1,N;h\neq
k}d^{3}\mathbf{r}_{h}\int_{U^{N}}\prod\limits_{j=1,N}d^{3}\mathbf{v}%
_{j}G_{i}(\mathbf{r}_{k},\mathbf{v}_{k}\mathbf{,}t)f_{N}(\mathbf{x},t)= \\
=\int_{U}d^{3}\mathbf{v}_{k}G_{i}(\mathbf{r}_{k},\mathbf{v}_{k}\mathbf{,}%
t)f_{1}(\mathbf{r}_{k},\mathbf{v}_{k}\mathbf{,}t)=Z_{i}(\mathbf{r}_{k}%
\mathbf{,}t).%
\end{array}%
\right.   \label{EQ-1a}
\end{equation}%
As suggested by classical statistical mechanics (CSM) \cite%
{Grad1958,Cercignani1969}, $G_{i}(\mathbf{r},\mathbf{v,}t)$ are
identified respectively with
\begin{equation}
G_{i}(\mathbf{r},\mathbf{v,}t)=\mathbf{v,\rho }_{0}u^{2}/3
\label{EQ-1b}
\end{equation}%
[with $\mathbf{u\equiv v-V}(\mathbf{r,}t)$ the relative velocity] for $%
\mathbf{V}(\mathbf{r,}t)$ and $p_{1}(\mathbf{r,}t)$;

\item \textit{Axiom }\#4 (\textit{entropy moments}): $f_{N}(t)\equiv f_{N}(%
\mathbf{x}_{1},\mathbf{...,x}_{N},t)$ determines uniquely the
global fluid field $S_{T}(t).$ Again based on CSM, the
thermodynamic entropy $S_{T}(t)$ can be identified with the
Boltzmann-Shannon (BS) statistical entropy. For this reason,
consistent with Ref.\cite{Tessarotto2007a} we require that for all
$t\in I$:

\begin{equation}
S_{T}(t)=S(f_{1}(t)),  \label{entropy moment}
\end{equation}%
where $f_{1}(t)\equiv f_{1}(\mathbf{x}_{1},t)$ and
$f_{1}(\mathbf{x}_{1},t)$ is defined in terms of $f_{N}(t)$ by
means of Eq.(\ref{reduced s-body PDF}). Furthermore we impose also
that for arbitrary $N\in
\mathbb{N}
_{1}$ and $t\in I$
\begin{equation}
K_{N}^{2}S(f_{N})=S(f_{1})  \label{constraint on S(f_N))}
\end{equation}%
(\textit{entropy constraint}). Here, denoting by $\Gamma ^{N}$ the
product
phase-space $\Gamma ^{N}\equiv \prod\limits_{i=1,N}\Gamma _{1},$ with $%
\Gamma _{1}=\Omega \times U$, the BS entropy for the $N$-point PDF
$f_{N}$ is defined as
\begin{equation}
S(f_{N})=-\int\limits_{\Gamma ^{N}}d\mathbf{x}f_{N}\ln f_{N},
\label{BS entropy}
\end{equation}%
where $d\mathbf{x}=\prod\limits_{k=1,N}d\mathbf{r}_{k}d\mathbf{v}_{k}$ and $%
K_{N}^{2}$ are suitable constants independent $f_{N}$ to be
determined$;$

\item \textit{Axiom }\#5 (\textit{entropic principle}): for all $N$ $\in
\mathbb{N}
_{1},$ $f_{N}(\mathbf{x}_{1},\mathbf{...,x}_{N},t)$ satisfies the
principle
of entropy maximization requiring%
\begin{equation}
\delta S(f_{1})=0  \label{PEM}
\end{equation}%
(PEM variational principle \cite{Jaynes1957}]). The variational principle (%
\ref{PEM}) is imposed either solely subject to \textit{Axiom \#5a} (\textit{%
local} \textit{entropic principle}) \textit{at some initial time
}$t=t_{o}$ or to \textit{Axiom \#5b} (\textit{global}
\textit{entropic principle}) \textit{for all} $t\in I.$
\end{itemize}

Let us analyze the physical interpretation of the previous
assumptions.

First we notice that \#1,\#2,\#3 and \#4 follow from the
requirement that
the state of the fluid is \textit{solely} prescribed by the set of \textit{%
local} and \textit{global} fluid fields $\left\{ Z\right\} $. \ In
particular the locality of the fluid fields, together with the
assumption that they are defined everywhere in $\overline{\Omega
},$ implies manifestly the symmetry requirement (\ref{INVARIANCE})
(see \#1). In fact, the positions
$\mathbf{r}_{1},\mathbf{...,r}_{N}$ can be manifestly interchanged
arbitrarily among them (and similarly the velocity vectors $\mathbf{v}_{1},%
\mathbf{...,v}_{N})$ without affecting the determination of the
fluid fields. This justifies the definitions given above in terms
of the 1-point
PDF both for the local and global fluid fields [see Eqs.(\ref{EQ-1a}), (\ref%
{entropy moment}) and (\ref{constraint on S(f_N))})].

In a similar way, since by assumption the fluid fields cannot
depend on the level adopted for the statistical description of the
fluid, all moments
which define the fluid field must be independent of the choice of the $N$%
-point PDF. Indeed, for arbitrary $N\in
\mathbb{N}
_{1},$ it must be possible to represent the thermodynamic entropy
in terms of Boltzmann-Shannon entropy associated to the $N$-point
PDF $f_{N},$ as well to $f_{1}$. This implies, that besides the
position (\ref{EQ-1a})
invoked for the local fluid fields also the additional constraint (\ref%
{constraint on S(f_N))}) must be placed\textit{\ }on \textit{all
the BS entropies associated to multi-point PDFs}. This constraint
manifestly should hold identically (for all $t\in I$)$.$\

Finally, the hypothesis that the Boltzmann-Shannon entropy is
maximal (see \#5) implies the validity of the entropic principle
(\ref{PEM}). We stress
that, in principle, PEM can be assumed to hold either at the initial time $%
t_{o}$ or, more generally, for arbitrary $t\in I.$ The second
requirement is
consistent with the assumption of isentropic flow. In fact, the positions (%
\ref{entropy moment}) and (\ref{constraint on S(f_N))}) imply that
also the BS entropy must be constant (i.e., independent of time).
Hence, the requirement that it is maximal at some initial time
$t_{o}$ may not be at variance with the requirement placed by the
global entropic principle \#5b$.$

Basic issues are related to the, possibly non-unique,
determination of the appropriate statistical model $\left\{
f,\Gamma \right\} $. \ These concern in particular:

\begin{enumerate}
\item (PROBLEM \#1) the search of the (possible) minimum level ($N$) of the
statistical description to be adopted for $\left\{ f,\Gamma
\right\} ;$

\item (PROBLEM \#2) the determination of the time-evolution of the
multi-point PDFs $f_{N};$

\item (PROBLEM \#3) the determination of the initial and boundary conditions
for $f_{N}$.
\end{enumerate}

Regarding the first problem the following remarkable result
holds:\\

\noindent \textbf{THM.1 - Factorization theorem for} $f_{N}.$

\textit{Let us impose Axioms \#1-\#4 with \#5b. Then it follows
necessarily that:}

\textit{1) the variational constraint}%
\begin{equation}
\delta \left\{ K_{N}^{2}S(f_{N})-S(f_{1})\right\} =0
\label{variational entropy constraint}
\end{equation}%
\textit{must hold for all }$t\in I;$

\textit{2) for all }$N\in
\mathbb{N}
_{1},$\textit{\ the }$N$-point\textit{\ PDF} $f_{N}(\mathbf{x}_{1},\mathbf{%
...,x}_{N},t)$ \textit{is of the form:}
\begin{equation}
f_{N}(\mathbf{x}_{1},\mathbf{...,x}_{N},t)=\prod\limits_{i=1,N}f_{1}(\mathbf{%
x}_{1},t),  \label{factorization}
\end{equation}%
\textit{with} $f_{1}(\mathbf{x}_{1},t)$ \textit{denoting the
corresponding 1-point PDF defined by Eq.(\ref{reduced s-body
PDF}). Hence, it follows also
that for all }$s=1,N-1$\textit{:}%
\begin{equation}
f_{s}(\mathbf{x}_{1},\mathbf{...,x}_{s},t)=\int\limits_{U}d^{3}\mathbf{v}%
_{s+1}\ f_{s+1}(\mathbf{x}_{1},\mathbf{...,x}_{s+1},t).
\label{reduced s-body II case}
\end{equation}

\textit{3) the constant }$K_{N}^{2}$\textit{\ in
Eq.(\ref{constraint on S(f_N))}) reads}
\begin{equation}
K_{N}^{2}=N\mu (\Omega )^{N-1}.  \label{constant}
\end{equation}
\noindent PROOF First we notice that the entropy constraint
(\ref{constraint on S(f_N))}) together the global entropic
principle\ \#5b [i.e., the
requirement that Eq.(\ref{PEM}) holds for all $t\in I$] imply that, for all $%
N$ and for all $t\in I,$ also the variational constraint
(\ref{variational entropy constraint}) must be fulfilled. To prove
that the factorization property of the $N$-point PDF must hold for
all $t\in I$, let us consider
for illustration (and without loss of generality) the case $N=2.$ Denoting $%
f_{2}(\mathbf{x}_{1},\mathbf{x}_{2},t)\equiv f_{2}(1,2)$ and $f_{1}(\mathbf{x%
}_{1},t)\equiv f_{1}(1)$, Eq.(\ref{variational entropy
constraint}) delivers
for arbitrary variations $\mathbf{\delta }f_{1}(3)$:%
\begin{equation}
\int\limits_{\Gamma ^{3}}d\mathbf{x\delta }f_{1}(3)\left\{
f_{2}(1,2)\ln f_{2}(1,2)-f_{1}(1)f_{1}(2)\left[ \ln f_{1}(1)+\ln
f_{1}(2)\right] \right\} =0.
\end{equation}%
This implies\ necessarily that the factorization condition $%
f_{2}(1,2)=f_{1}(1)f_{1}(2)$ must hold identically in $\Gamma
^{2}\times I.$
The proof can easily be extended to arbitrary $N>2,$ yielding Eq.(\ref%
{factorization}). In turn, thanks to Eq.(\ref{factorization}), equations (%
\ref{reduced s-body II case}) and (\ref{constant}) immediately
follow, respectively from Eqs.(\ref{reduced s-body PDF}) and
(\ref{constraint on S(f_N))}). Q.E.D.

We remark that in principle THM.1 can be generalized by requiring
that PEM holds only at the initial time $t_{o}\in I$ (Axiom
\#5a)$.$ Nevertheless, in this case the constraint
(\ref{constraint on S(f_N))}) only warrants that the factorization
condition (\ref{reduced s-body II case}) holds at the initial time
$t_{o},$ \textit{unless} the form of the statistical (Liouville)
equations holding for the $s$-point velocity PDFs is explicitly
prescribed as done in Ref. \cite{Tessarotto2010c}.

Invoking, however, the \textit{validity of Axiom \#5b and
consequently of THM.1, the statistical model }$\left\{ f,\Gamma
\right\} $\textit{\ can be
identified with the IKT statistical model for the 1-point PDF \cite%
{Ellero2005,Tessarotto2006,Tessarotto2007a,Tessarotto2008z}.}

\section{IKT for multi-point PDFs}

The construction of multi-point PDFs is a problem of "practical"
interest in experimental/numerical research in fluid dynamics,
usually adopted for the statistical analysis of turbulent fluids.
In fact, they can be experimentally measured in terms of velocity
differences between different fluid elements.

Let us assume, for definiteness, that
$f_{1}(\mathbf{x}_{i}\mathbf{,}t)$ is the $1-$point PDF which is
particular solution of the Liouville equation [or \textit{inverse
kinetic equation }(IKE)] provided by IKT \cite{Ellero2005}.
Then, denoting $f_{1}(i)\equiv f_{1}(\mathbf{x}_{i}\mathbf{,}%
t)$ (for $i=1,s$) the same PDF evaluated at the states
$\mathbf{x}_{i}\equiv \left( \mathbf{r}_{i},\mathbf{v}_{i}\right)
$ (for $i=1,s$)$,$ the $s-$point PDF is the probability density
\begin{equation}
f_{s}(1,2,..s)\equiv \prod\limits_{i=1,s}f_{1}(i),
\end{equation}%
defined in the product phase-space $\Gamma ^{s}\equiv
\prod\limits_{i=1,s}\Gamma,$ The statistical equation advancing in time $%
f_{s}$ follows trivially from the Liouville equation for the 1-point PDF {%
see \cite{Ellero2005}). In fact, denoting for $i=1,s$ by $\mathbf{F}%
(i)\equiv \mathbf{F}(\mathbf{x}_{i},t;f_{1})$ the 1-point
mean-field force per unit mass acting on the }$i$-th particle
(with{\ state $\mathbf{x}_{i})$ [defined in Refs.
\cite{Ellero2005} and \cite{Tessarotto2006}] and introducing the
$s-$point Liouville operator
\begin{equation}
L_{s}(1,..,s)\equiv \frac{\partial }{\partial
t}+\sum\limits_{i=1,s}\left[
\mathbf{v}_{i}\mathbf{\cdot }\frac{\partial }{\partial \mathbf{r}_{i}}+\frac{%
\partial }{\partial \mathbf{r}_{i}}\cdot \left\{ \mathbf{F}_{i}(i)\right\} %
\right] ,  \label{s-point Liouville op}
\end{equation}%
it follows that $f_{s}(1,2,..s)$ satisfies identically the
$s-$point
Liouville equation%
\begin{equation}
L_{s}(1,..,s)f_{s}(1,2,..s)=0.  \label{s-point Liouville equation}
\end{equation}%
}

\subsection{Explicit evaluation of 2-point velocity PDFs}

In terms of the 2-point PDF, $f_{2}(1,2)$, a number of reduced
probability densities can be defined in suitable subspaces of
$\Gamma ^{2}$. To introduce them explicitly let us first introduce
the transformation to the center of mass coordinates of the two
point-particles with states $\left(
\mathbf{r}_{i},\mathbf{v}_{i}\right) $ (for $i=1,2$)
\begin{equation}
\left\{
\mathbf{r}_{1},\mathbf{v}_{1},\mathbf{r}_{2},\mathbf{v}_{2}\right\}
\rightarrow \left\{ \mathbf{r,R,v,V}\right\}
\end{equation}%
[here $\mathbf{r=\frac{\mathbf{r}_{1}-\mathbf{r}_{2}}{2},R=}\frac{\mathbf{r}%
_{1}+\mathbf{r}_{2}}{2};$ furthermore, $\mathbf{v,V}$ can be identified with$%
\ \mathbf{v=\mathbf{v}_{1}-\mathbf{v}_{2}}$ and $\mathbf{V=v}_{1}+\mathbf{v}%
_{2}$]. Then, these are respectively:

1) the\textit{\ local }(in configuration space)\textit{\
velocity-difference 2-point PDF
}$g_{2}(\mathbf{r}_{1},\mathbf{r}_{2},\mathbf{v},t)$ defined in
the phase-space $\Omega ^{2}\times U$ and obtained integrating the
2-point velocity PDF w.r. to the mean velocity $\mathbf{V}$
\begin{equation}
\left.
\begin{array}{c}
g_{2}(\mathbf{r}_{1},\mathbf{r}_{2},\mathbf{v},t)=\int_{U}d^{3}\mathbf{V}%
f_{2}(1,2)\equiv  \\
\equiv \int d^{3}\mathbf{V}f_{1}(\mathbf{r}_{1}\mathbf{,v+V,}t))f_{1}(%
\mathbf{r}_{2},\mathbf{V-v,}t);%
\end{array}%
\right.   \label{g2}
\end{equation}

2) the \textit{velocity-difference 2-point PDF} $\widehat{f}_{2}(\mathbf{r},%
\mathbf{v,}t)$ defined in $\Gamma _{1}=\Omega \times U$ and
obtained integrating also on the center-of-mass position vector
$\mathbf{R.}$ Thus
denoting by%
\begin{equation}
\left. \left\langle \cdot \right\rangle _{\mathbf{R,}\Omega
}=\frac{1}{\mu (\Omega )}\int_{\Omega }d^{3}\mathbf{R}\cdot
\right. \label{averaging operator}
\end{equation}

the configuration-space average operator acting on the center of
mass
coordinates $\mathbf{R,}$ there it follows%
\begin{equation}
\widehat{f}_{2}(\mathbf{r},\mathbf{v,}t)=\left\langle g_{2}(\mathbf{r+R},%
\mathbf{R-r},\mathbf{v},t)\right\rangle _{\mathbf{R,}\Omega }.
\label{2-point velocity difference PDF}
\end{equation}%
In particular, in the case of a Gaussian PDF \cite{Tessarotto2010c}, Eq.(\ref%
{g2}) delivers again a Gaussian-type PDF%
\begin{equation}
g_{2}(\mathbf{r}_{1},\mathbf{r}_{2},\mathbf{v},t)=\frac{1}{\pi
^{3/2}v_{th}^{3}}\exp \left\{ -\frac{\left\Vert \mathbf{v-}\frac{\mathbf{V}%
(1)-\mathbf{V}(2)}{2}\right\Vert ^{2}}{v_{th}^{2}}\right\} ,
\label{GAUSSIAN 2-POINT PDF}
\end{equation}%
where $\mathbf{V}(i)\equiv \mathbf{V}(\mathbf{r}_{i},t),$ $%
v_{th,p}^{2}(i)=v_{th,p}^{2}(\mathbf{r}_{i},t)$ and $v_{th}^{2}$
denotes

\begin{equation}
v_{th}^{2}=\frac{v_{th,p}^{2}(1)+v_{th,p}^{2}(2)}{4}.
\end{equation}

In a similar way it is possible to obtain explicit representations
for the following additional 2-point PDFs: \textit{\ }

\begin{enumerate}
\item \textit{the velocity-difference 2-point PDF for parallel velocity
increments}$.$\textit{\ }Introducing the representations $\mathbf{v}=\mathbf{%
n}v$ and $\mathbf{r}=\mathbf{n}r,$ $\mathbf{n}$ denoting a unit vector, $%
\widehat{f}_{2\parallel }(r,v\mathbf{,}t)$ can be simply defined
as the
solid-angle average%
\begin{equation}
\widehat{f}_{2\parallel }(r,v\mathbf{,}t)=\int d\Omega (\mathbf{n})\widehat{f%
}_{2}(\mathbf{r=n}r,\mathbf{v}=\mathbf{\mathbf{n}}v\mathbf{,}t);
\label{2-point velocity-difference PDF for parallel}
\end{equation}

\item \textit{the velocity-difference 2-point PDF for perpendicular velocity
increments}$.$\textit{\ } Introducing, instead, the representations $\mathbf{%
v}=\mathbf{n}v$ and $\mathbf{r}=\mathbf{n\times b}r,$ $\mathbf{n}$ and $%
\mathbf{b}$ denoting two independent unit vectors, $\widehat{f}_{2\perp }(r,v%
\mathbf{,}t)$ can be defined as the double-solid-angle average%
\begin{align}
& \left. \widehat{f}_{2\perp }(r,v\mathbf{,}t)=\int d\Omega
(\mathbf{n})\int d\Omega (\mathbf{b})\right.
\label{2-point velocity-difference PDF for perpendicular} \\
& \left. \widehat{f}_{2}(\mathbf{r=n\times b}r,\mathbf{v}=\mathbf{\mathbf{n}}%
v\mathbf{,}t).\right.   \notag
\end{align}

An interesting property which emerges from these results is that
in all
cases indicated above [i.e., Eqs.(\ref{2-point velocity difference PDF}),(%
\ref{2-point velocity-difference PDF for parallel}) and
(\ref{2-point velocity-difference PDF for perpendicular})] the
definition of $g_{2}$ given
above [Eq.(\ref{g2})] implies that non-Gaussian features, respectively in $%
\widehat{f}_{2},\widehat{f}_{2\parallel }$ and
$\widehat{f}_{2\perp },$ may arise even if the $1-$point PDF is
Gaussian. This occurs due to velocity and
pressure fluctuations occurring between different spatial positions $\mathbf{%
r}_{1}$ and $\mathbf{r}_{2}$. More generally, however, we can
infer that,
due to the constraint here imposed on the 1-point PDF%
\begin{equation}
\left\langle f_{1}(t)\right\rangle _{\mathbf{r},\Omega }=\widehat{f}%
_{1}^{(freq)}(t)  \label{constraint on f1 (at initial time)}
\end{equation}%
[where $\left\langle \cdot \right\rangle _{\mathbf{r},\Omega }$ it
the averaging operator $\left\langle \bullet \right\rangle
_{\mathbf{r,}\Omega }\equiv \frac{1}{\mu (\Omega
)}\int\limits_{\Omega }d^{3}\mathbf{r}_{o}$ acting on of a
function $F(\mathbf{x},t)$], it is obvious that, if the fluid
velocity $\mathbf{V}(\mathbf{r},t)$ is bounded in the domain $\overline{%
\Omega },$\textit{\ the same 1-point PDF, and hence the 2-point
PDFs, cannot be Gaussian distributions}.
\end{enumerate}

\subsection{Statistical evolution equation for the velocity-difference
2-point PDF}

From the $2-$point IKE (\ref{s-point Liouville equation})
(obtained in the case $s=2$) it is immediate to obtain the
corresponding evolution equation for the reduced PDFs indicated
above. For example, the velocity-difference 2-point PDF
$\widehat{f}_{2}$ satisfies the equation
\begin{equation}
\frac{\partial \widehat{f}_{2}}{\partial t}+\mathbf{v\cdot }\frac{\partial }{%
\partial \mathbf{r}}\widehat{f}_{2}=-\frac{\partial }{\partial \mathbf{v}}%
\cdot \mathbf{D}  \label{F-P-1-a}
\end{equation}%
where $\mathbf{D}$ is the diffusion vector
\begin{equation}
\mathbf{D}=\int d^{3}\mathbf{V}\left\langle \frac{\mathbf{F}_{1}(1)-\mathbf{F%
}_{2}(2)}{2}f_{2}(1,2)\right\rangle _{\mathbf{R,\Omega }}.
\label{FP-1}
\end{equation}%
It follows, in particular, that in the case of a Gaussian 1-point
PDF this
equation reduces to the Fokker-Planck equation%
\begin{equation}
\frac{\partial \widehat{f}_{2}}{\partial t}+\mathbf{v\cdot }\frac{\partial }{%
\partial \mathbf{r}}\widehat{f}_{2}=-\frac{\partial }{\partial \mathbf{v}}%
\cdot \widehat{\mathbf{D}}  \label{F-P-2-a}
\end{equation}%
where the Fokker-Planck diffusion vector $\widehat{\mathbf{D}}$
reads
\begin{equation}
\widehat{\mathbf{D}}=\left\langle \mathbf{F}^{(T)}g_{2}(\mathbf{r+R},\mathbf{%
R-r},\mathbf{v},t)\right\rangle _{\mathbf{R,\Omega }} \label{FP-2}
\end{equation}%
and the vector field $\mathbf{F}_{1}^{(T)}\equiv \mathbf{F}_{1}^{(T)}(%
\mathbf{r}_{1}\mathbf{,r}_{2},\mathbf{V},t\mathbf{;}f_{M})$ is
reported in Ref. \cite{Ellero2005}. It follows that both equations
are manifestly \textit{non-Markovian }as a consequence of the
non-local dependencies
arising (in both cases) in the Fokker-Planck coefficients $\mathbf{D}$ and $%
\widehat{\mathbf{D}}$.

An interesting issue is here provided by the comparison with the
statistical
formulation developed by Peinke and coworkers \cite%
{Naert1997,Friedrich1999,Luck1999,Renner2001,Renner2002}. Their
approach, based on the statistical analysis of experimental
observations, indicates that in case of stationary and homogeneous
turbulence both the 2-point PDFs for parallel and velocity
increments obey stationary Fokker-Planck
equations. In particular, according to experimental evidence \cite%
{Renner2001,Renner2002} a reasonable agreement with a Markovian
approximation for Eq.(\ref{F-P-2-a}) - at least in some limited
subset of parameter space- is suggested. Our theory implies,
however, that a breakdown of the Markovian property should be
expected due to non-local contributions
appearing in the previous statistical equations (\ref{F-P-1-a}) and (\ref%
{F-P-2-a}).

\section{Conclusions}

In this paper we have shown that the multi-point PDFs used in
customary phenomenological approaches to turbulence can be
explicitly evaluated in terms of the 1-point velocity PDF
($f_{1}$) determined in the framework on
the IKT-statistical model \cite%
{Ellero2005,Tessarotto2006,Tessarotto2007a,Tessarotto2008z,Tessarotto2010c}.

The starting point is provided by THM.1, which shows that under
suitable hypotheses the multi-point PDF $f_{N}$ is necessarily
factorized in terms of the 1-point PDF $f_{1}.$ The requirements
here imposed include, in particular, the assumption that $\left\{
f_{N},\Gamma ^{N}\right\} $ is a complete statistical model, i.e.,
that in terms of the multi-point PDF the complete set of fluid
fields (defining the fluid state) can be represented by means of
suitable velocity and phase-space moments [see Axioms \#1-\#6].
Then, provided:

A) the entropy constraint\textit{\ }(\ref{constraint on S(f_N))})
is invoked ((Axiom \#4);

B) the validity of PEM is imposed \textit{at all times} $t\in I$
(Axiom \#5b);

the factorization condition (\ref{factorization}) for $f_{N}$ in
terms of the 1-point PDF $f_{1}$ necessarily follows.

As a result, in validity of the previous requirements, the
statistical model for NS fluid can be identified with the
IKT-statistical model $\left\{
f_{1},\Gamma _{1}\right\} $ earlier developed \cite%
{Ellero2005,Tessarotto2006,Tessarotto2007a,Tessarotto2008z} and
based on the 1-point PDF $f_{1}$. The theory has important
consequences:

\begin{enumerate}
\item arbitrary multi-point PDFs can be uniquely represented in terms of the
1-point PDF characterizing the IKT-statistical model $\left\{
f_{1},\Gamma _{1}\right\} $;

\item the time evolution of the multi-point PDFs is uniquely determined by $%
\left\{ f_{1},\Gamma _{1}\right\} ;$

\item the \textit{theoretical prediction} of multipoint PDFs is actually
possible.

\item qualitative properties of the multi-point PDFs can be investigated. As
a particular case, the example of a Gaussian 1-point PDF has been
pointed out. \
\end{enumerate}

In the IKT-statistical model the statistical equation advancing in
time the $1-$point PDF $f_{1}$ coincides with the Liouville
equation. As a consequence, its explicit evaluation is actually
made possible \cite{Tessarotto2010c}. In particular, as shown in
Ref. \cite{Tessarotto2010d}, in the presence of HIST the 1-point
PDF necessarily coincides with a Gaussian distribution. Thanks to
the factorization theorem (THM.1) this implies that also the
multi-point velocity PDFs are uniquely determined. As result, as
indicated in Section 3 (see subsection 3.1), two-point PDFs
relevant for the phenomenological description of hydrodynamic
turbulence can be explicitly determined.

\section{Appendix: INSE Problem}

The fluid equations for a NS fluid are the so-called
incompressible Navier-Stokes\ equations (INSE) for the fluid
fields $\left\{ Z\right\}
\equiv \left\{ \rho _{0},\mathbf{V}(\mathbf{r,}t),p_{1}(\mathbf{r,}%
t),S_{T}(t)\right\} :$
\begin{eqnarray}
\rho  &=&\rho _{o},  \label{1b} \\
\nabla \cdot \mathbf{V} &=&0,  \label{1ba} \\
N\mathbf{V} &=&0,  \label{1bbb} \\
\frac{\partial }{\partial t}S_{T} &=&0,  \label{1bb} \\
Z(\mathbf{r,}t_{o}) &\mathbf{=}&Z_{o}(\mathbf{r}),  \label{1ca} \\
\left. Z(\mathbf{r,}t)\right\vert _{\partial \Omega }
&\mathbf{=}&\left. Z_{w}(\mathbf{r,}t)\right\vert _{\partial
\Omega },  \label{1c}
\end{eqnarray}

Eqs. (\ref{1b})- (\ref{1c}) denote respectively the \textit{%
incompressibility, isochoricity, Navier-Stokes and constant
thermodynamic
entropy equations} and the initial and Dirichlet boundary conditions for $%
\left\{ Z\right\} ,$ with $\left\{ Z_{o}(\mathbf{r})\right\} $ and
$\left\{ \left. Z_{w}(\mathbf{r,}t)\right\vert _{\partial \Omega
}\right\} $ suitably prescribed initial and boundary-value fluid
fields, defined respectively at the initial time $t=t_{o}$ and on
the boundary $\partial \Omega,$ In particular, this means that
they are are required to be at least continuous
in all points of the closed set $\overline{\Omega }\times I$, with $%
\overline{\Omega }=\Omega \cup \partial \Omega $ closure of
$\Omega ${$.$ \ In the remainder we shall require that: }

\begin{enumerate}
\item {$\Omega ${\ (\textit{configuration domain}) is a bounded subset of
the Euclidean space }$E^{3}$ on }$%
\mathbb{R}
${$^{3};$}

\item $I$ (\textit{time axis}) is identified, when appropriate, either with
a bounded interval, \textit{i.e.}, $I${$=$}$\left]
{t_{0},t_{1}}\right[ \subseteq
\mathbb{R}
,$ or with the real axis $%
\mathbb{R}
$;

\item in the open set $\Omega \times ${$I$} the functions $\left\{ Z\right\}
,$ are assumed to be solutions of Eqs.(\ref{1ba})-(\ref{1bb})
subject, while
in $\overline{\Omega }\times ${$I$} they satisfy the whole set of Eqs. (\ref%
{1b})-(\ref{1c}).\textit{\ }In particular: {Eqs. (\ref{1b})-
(\ref{1c}) define the \textit{initial-boundary value INSE problem,
}}

\item by assumption, the fluid fields are \textit{strong solutions} of the
fluid equations. Hence Eqs. (\ref{1b})- (\ref{1c}) are required to
define a well-posed problem with unique strong solution defined
everywhere in $\Omega \times ${$I$}.
\end{enumerate}

Here the notation as follows. $N$ is the \textit{NS nonlinear
operator}
\begin{equation}
N\mathbf{V}=\frac{D}{Dt}\mathbf{V}-\mathbf{F}_{H},  \label{NS
operator}
\end{equation}%
with $\frac{D}{Dt}\mathbf{V}$ and\textbf{\ }$\mathbf{F}_{H}$
denoting
respectively the \textit{Lagrangian fluid acceleration} and the \textit{%
total force} \textit{per unit mass}
\begin{eqnarray}
&&\left. \frac{D}{Dt}\mathbf{V}=\frac{\partial }{\partial t}\mathbf{V}(%
\mathbf{r,}t)+\mathbf{V}(\mathbf{r,}t)\cdot \nabla \mathbf{V(\mathbf{r}},t%
\mathbf{),}\right.  \\
&&\left. \mathbf{F}_{H}\equiv \mathbf{-}\frac{1}{\rho _{o}}\nabla p(\mathbf{%
r,}t)+\frac{1}{\rho _{o}}\mathbf{f}(\mathbf{r,}t)+\upsilon \nabla ^{2}%
\mathbf{V(\mathbf{r,}}t\mathbf{),}\right.   \label{2c}
\end{eqnarray}%
while $\rho _{o}>0$ and $\nu >0$ are the \textit{constant mass
density} and the constant \textit{kinematic viscosity}. In
particular, $\mathbf{f}$ is the \textit{volume force density}
acting on the fluid, namely which is
assumed of the form%
\begin{equation}
\mathbf{f=-\nabla }\phi
((\mathbf{r,}t)+\mathbf{f}_{R}(\mathbf{r,}t), \label{potential}
\end{equation}%
$\phi ((\mathbf{r,}t)$ being a suitable scalar potential, so that
the first two force terms [in Eq.(\ref{2c})] can be represented as
\begin{equation}
-\nabla p(\mathbf{r,}t)+\mathbf{f}(\mathbf{r,}t)=-\nabla p_{1}(\mathbf{r,}t)+%
\mathbf{f}_{R}(\mathbf{r,}t),
\end{equation}%
with $p_{1}(\mathbf{r,}t)$ defined by Eq.(\ref{p_1}) denoting the
kinetic pressure. As a consequence the fluid pressure necessarily
satisfies the
\textit{Poisson equation}%
\begin{equation}
\nabla ^{2}p(\mathbf{r,}t)=S(\mathbf{r,}t),  \label{Poisson}
\end{equation}%
where the source term $S$ reads
\begin{equation}
S(\mathbf{r,}t)=-\rho _{o}\nabla \cdot \left( \mathbf{V}\cdot \nabla \mathbf{%
V}\right) +\nabla \cdot \mathbf{f}.
\end{equation}


\bigskip
\section{Acknowledgments}

Work developed in cooperation with the CMFD Team, Consortium for
Magneto-fluid-dynamics (Trieste University, Trieste, Italy).
Research partially performed in the framework of the GDRE (Groupe
de Recherche Europeenne) GAMAS.

\end{document}